\documentclass{article}
\usepackage{spconf,graphicx}
\usepackage{amsmath,amssymb,amsfonts}
\usepackage{multirow}
\usepackage{booktabs}
\usepackage{xcolor,colortbl}
\usepackage{dblfloatfix}
\usepackage{caption}
\usepackage[noadjust]{cite}
\usepackage{hyperref}
\usepackage{fancyhdr}

\usepackage{multirow}
\usepackage{tabularx}

\newcommand{\HH}[1]{\ignorespaces}
\newcommand{\PG}[1]{\ignorespaces}
\newcommand{\KL}[1]{\ignorespaces}

\newcommand{\cut}[1]{}

\title{Self-supervised learning for infant cry analysis}

\name{Arsenii Gorin$^{\star}$, Cem Subakan$^{\natural\sharp \flat}$, Sajjad Abdoli$^{\star}$, Junhao Wang$^{\star}$, Samantha Latremouille$^{\star}$, Charles Onu$^{\star\flat}$}

\address{$^{\star}$Ubenwa Health, Montr\'eal, Canada \; $^{\natural}$Universit\'e Laval, Qu\'ebec City, Canada \\ $^{\sharp}$Concordia University, Montr\'eal, Canada \; $^\flat$Mila-Qu\'ebec AI Institute, Montr\'eal, Canada  
			     }

\newlength{\bibitemsep}
\newlength{\bibparskip}
\let\oldthebibliography\thebibliography
\renewcommand\thebibliography[1]{%
  \oldthebibliography{#1}%
  \setlength{\parskip}{\bibitemsep}%
  \setlength{\itemsep}{\bibparskip}%
}

\makeatother

\fancypagestyle{FirstPage}{
\lfoot{\scriptsize © 2023 IEEE.  Personal use of this material is permitted. Permission from IEEE must be obtained for all other uses, in any current or future media, including reprinting/republishing this material for advertising or promotional purposes, creating new collective works, for resale or redistribution to servers or lists, or reuse of any copyrighted component of this work in other works.} 
}

\begin{document}

\maketitle

\copyrightnotice{\copyright yy}

\ninept

\newcommand{\cem}[1]{#1} 

\begin{abstract}

In this paper, we explore self-supervised learning (SSL) for analyzing a first-of-its-kind database of cry recordings containing clinical indications of more than a thousand newborns. Specifically, we target cry-based detection of neurological injury as well as identification of cry triggers such as pain, hunger, and discomfort. 
Annotating a large database in the medical setting is expensive and time-consuming, typically requiring the collaboration of several experts over years. Leveraging large amounts of unlabeled audio data to learn useful representations can lower the cost of building robust models and, ultimately, clinical solutions.
In this work, we experiment with self-supervised pre-training of a convolutional neural network on large audio datasets. 
We show that pre-training with SSL contrastive loss (SimCLR) performs significantly better than supervised pre-training for both neuro injury and cry triggers. In addition, we demonstrate further performance gains through SSL-based domain adaptation using unlabeled infant cries. We also show that using such SSL-based pre-training for adaptation to cry sounds decreases the need for labeled data of the overall system.

\end{abstract}

\begin{keywords}
Self-Supervised Learning, Infant Cry Classification, Audio Classification, Transfer Learning, Domain Adaptation
\end{keywords}

\section{Introduction}

\thispagestyle{FirstPage}
\label{sec:intro}

Crying is the primary means by which babies communicate with the world.
Researchers have been interested in infant cry analysis since the early 1960s~\cite{wasz1964identification}. 
Cry characteristics may help us to understand basic baby needs (hunger, pain, etc.) and, more importantly, can be analyzed for the early and non-invasive detection of various diseases~\cite{ji2021review}. 
 For example, clinical research has reported that certain infant cry characteristics are correlated with birth asphyxia~\cite{michelsson1977pain}. This multi-causal condition frequently leads to severe health problems, including neurological injury and even death. 
Various methods based on signal processing~\cite{liu2019infant}, statistical modeling~\cite{felipe2019identification,parga2020defining} and deep learning~\cite{onu2019neural,al2019vcmnet,ozseven2023infant,patil2022constant} have been explored for finding clinical and other insights using cry recordings. \\ 
\indent One of the main challenges in baby cry analysis is data acquisition. Today, cry sounds are not part of routine medical records, so obtaining a database requires targeted efforts such as a clinical study. These are expensive to conduct and typically require the collaboration of several hospital staff over the years. Most machine learning (ML) research on pathology detection from cry sounds was done using the Baby Chillanto~\cite{reyes2004system} database, which contains only six patients diagnosed with birth asphyxia.

From an ML problem point of view, cry classification is analogous to general audio classification, where deep convolutional neural networks (CNNs) have excelled as the state-of-the-art. 
Recently,~\cite{10.1109/TASLP.2020.3030497} demonstrated that Pre-trained Audio Neural Networks (PANNs) - large CNNs pre-trained on generic audio - transferred to a wide range of audio pattern recognition tasks outperformed several previous state-of-the-art systems. Since then, PANNs have been widely adopted for various audio tasks, including emotion recognition from speech~\cite{triantafyllopoulos2021role} and COVID-19 detection from cough~\cite{casanova2021transfer}.

Another popular paradigm in audio classification state-of-the-art is self-supervised learning (SSL) - a method to obtain high-quality representations by training on unlabeled data. SSL has revolutionized the fields of Natural Language Processing and Computer Vision and is currently widely adopted in audio processing~\cite{mohamed2022self}. 
A neural network (encoder) pre-trained with SSL can be seen as a non-linear mapping of an audio sequence to a hidden representation - an embedding. 
The embeddings can be used as input to a classifier trained on a specific task with a supervised objective (using labeled data and conventional cross-entropy loss).
This approach is common for benchmarking various SSL models on multiple diverse audio tasks~\cite{yang2021superb, wang2022towards}. 
Recently, a similarity-based contrastive learning method called SimCLR introduced in Computer Vision~\cite{chen2020simple,chen2020big} demonstrated good performance in multiple audio tasks~\cite{wang2022towards,wang2022learning}, including music  analysis~\cite{spijkervet2021contrastive,mccallum2022supervised}.
SimCLR maximizes the similarity between modified (distorted) views of the same object. For audio, such distortion can be done, for example, by mixing random audio samples~\cite{wang2022towards}, spectrogram masking~\cite{park2019specaugment} in~\cite{wang2022learning}, or/and reverberation, pitch shifting, etc~\cite{spijkervet2021contrastive}. 

In this paper, we experiment with PANNs using both supervised and self-supervised pre-training to learn representations for two  downstream tasks. The first task is classifying brain injury (resulting from birth asphyxia), and the second is predicting cry triggers (pain, hunger, discomfort). The methods are tested on a unique clinical database of newborn cries collected by Ubenwa Health in collaboration with hospitals across three countries~\cite{onu1711ubenwa}. 

In addition, we evaluate the impact of SimCLR-based adaptation of PANNs using unlabeled cries inspired by self-supervised domain adaptation in Speech~\cite{chen2021self} and Natural Language Processing~\cite{gururangan2020don}. 
It should be noted that speech and audio SSL state-of-the-art frequently uses transformers instead of CNNs and relies on different learning objectives~\cite{mohamed2022self}. However, our preliminary experiments with some popular pre-trained speech and audio transformers (specifically, Wav2Vec2.0~\cite{baevski2020wav2vec}, HuBERT~\cite{hsu2021hubert}, WavLM~\cite{chen2022wavlm} and  SSAST~\cite{gong2022ssast}) have not shown sufficient improvements but generally required many parameters to be adapted and hyperparameters tuned. We, therefore, focus on CNN and SimCLR, which demonstrated a good balance of accuracy and adaptation complexity.


\section{Methodology}
\label{sec:method}

We illustrate our 3-stage approach in Figure~\ref{fig:method}, indicating the SSL pipeline and datasets used at different stages. In the rest of this section, we provide details on each stage.
\newcommand{\SSLtwo}{SSL cry adaptation}
\begin{figure}[h]
    \centering
    \includegraphics[width=0.5\textwidth]{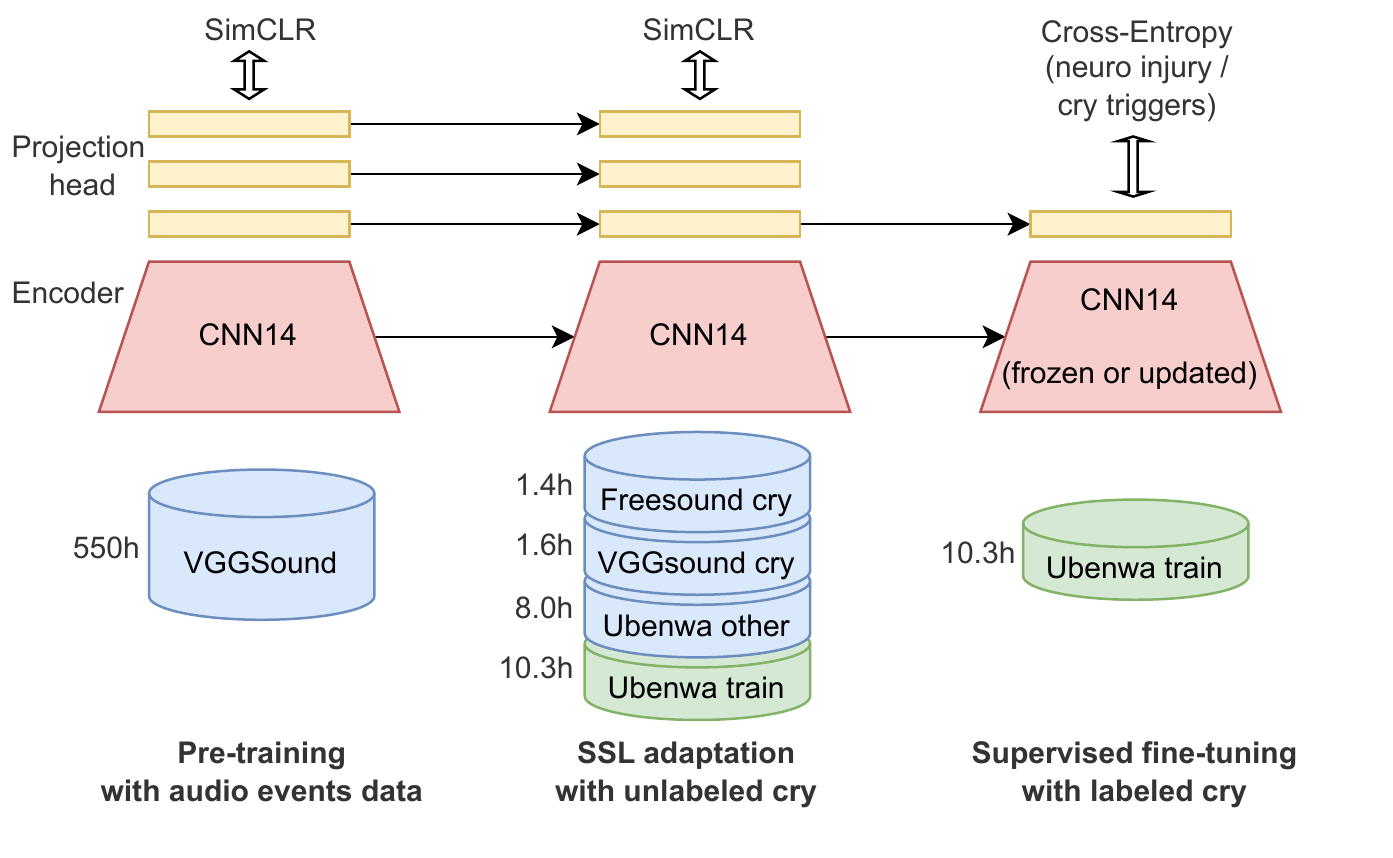}
    \caption{Summary of the proposed \cem{SSL-based trainining} pipeline. \cem{\textbf{(Left)} First the CNN14 backbone is pre-trained on the VGGSound dataset using SimCLR. \textbf{(Middle)} The CNN14 backbone is further pre-trained via SSL using cry-specific datasets. We denote this stage \emph{\SSLtwo}. \textbf{(Right)} The model is finally trained with supervision on a labeled cry dataset.}}
    \label{fig:method}
\end{figure}

\subsection{Encoder Architecture}

As the backbone encoder, we adopt CNN14 introduced in~\cite{10.1109/TASLP.2020.3030497}. The encoder is pre-trained on the VGGSound database as in ~\cite{wang2022learning}.

There are a few minor differences between the CNN14 proposed in~\cite{10.1109/TASLP.2020.3030497} and the models that we adopted from~\cite{wang2022learning}. First,~\cite{wang2022learning} uses VGGSound~\cite{Chen20} instead of AudioSet~\cite{45857} for pre-training. VGGSound is an automatically curated collection of 550 hours of YouTube recordings, which is about ten times smaller than AudioSet but with a single label per audio recording and a lower noise level.
Second,~\cite{wang2022learning} used more log-Mel filterbanks (80 vs 64) and shorter audio chunks in training (4 seconds vs 10). 
The network overall has about 80 million parameters. 

\subsection{Pre-Training}

We compare two identical CNN14 models pre-trained in a supervised and self-supervised manner on VGGSound. SimCLR pre-training relies on an additional projection head - a three-layer multi-layer perceptron with 2048 hidden units and a bottleneck layer with 512 hidden units.
In both supervised and self-supervised initial pre-training, the model is updated with stochastic gradient descent using mini-batches of 32 four-second chunks randomly cut from VGGSound clips. When extracting embeddings of validation and test data, the model processes arbitrary-length audio sequences using global temporal pooling. 

\subsection{Supervised Fine-Tuning and Evaluation}
\label{sec:protocol}

To study various aspects of pre-trained models, we conduct three evaluations. 
First, we use {\it linear probing}, where the frozen encoder  extracts features for a linear classifier. 
Second, in addition to learning the linear classifier, we also update statistics of batch normalization parameters of the encoder while still keeping other parameters frozen. The primary motivation is to compensate for the difference between pre-training and target data characteristics~\cite{frankle2020training,yazdanpanah2022revisiting}. This also allows us to understand what portion of improvement obtained by SSL fine-tuning with cry data may be attributed to a simple update of normalization parameters occurring naturally during SimCLR. 
Finally, aiming to improve classification results further, we perform {\it end-to-end fine-tuning}, where the encoder parameters are optimized jointly with the classifier on target tasks.

The supervised training is done for 50 epochs in all three settings, and the model with the best validation score is selected. We use weighted random sampling to balance the distribution of classes during training. 
We use Adam optimizer~\cite{kingma2014adam} with a learning rate reduced two times if validation loss does not improve for three epochs. For end-to-end supervised fine-tuning, a much smaller and separately tuned learning rate is used for the encoder. Also, the encoder learning rate is linearly increased from zero to target one over the first ten epochs. 
In all experiments, the learning rates of the classifier and encoder are optimized using grid search. For the model with the best validation score, we repeat the experiment 10 times and report the mean area under the receiver operating characteristic curve (AUC) along with standard error. For multi-class classification (triggers), the macro averaged AUC is computed using a one-versus-rest approach.

Similar to~\cite{chen2020big}, in our preliminary studies, we found that keeping one layer of projection head of the SSL pre-trained model leads to slightly better results. Therefore, we always transfer from layer 1 of the projection head.

\subsection{Self-Supervised Cry Adaptation}

To improve the quality of SSL representations, we further explore a second stage of self-supervised domain adaptation. Our goal is to adapt the encoder from the domain of general audio sounds to the domain of cry sounds, using unlabeled cry data (middle column of Figure \ref{fig:method}). This SSL adaptation is done by reusing the encoder and the projector from CNN14 trained on VGGSound and running 100 more epochs of SimCLR using infant cry data with the same learning rate and schedule as the initial pre-training. The only difference is that we use batch size 200 for faster training and because larger batches performed better for SimCLR in the literature~\cite{spijkervet2021contrastive}. Notably, we did not find a significant difference trying to improve the initial SSL pre-trained model of~\cite{wang2022learning} by using more training epochs and large batches without cry sounds.

Similar to the initial SSL pre-trained models, the cry-adapted ones are evaluated with linear probing and end-to-end settings described in the previous section. \cut{We expect linear probing of the cry-adapted models to perform significantly better than the non-adapted ones. This is perhaps less obvious for end-to-end fine-tuning as PANN may forget some generic knowledge useful for supervised fine-tuning.}

\section{Experimental setup}

\label{sec:exp_setup}

\subsection{Dataset Description}

This study is based on a subset of a larger Ubenwa newborn cry clinical database collected from five hospitals in Nigeria, Brazil, and Canada since 2020~\cite{onu1711ubenwa}. For most infants, one 
 recording is done after birth and one before discharge. A neurological exam was conducted on all infants, and the level of neuro injury was recorded using a four-scale measure called Sarnat score~\cite{sarnat1976neonatal}. For classification, we categorize the recordings into two groups: normal (no neuro injury) and injured (mild, moderate, or severe injury). 
We further split the data into train, validation, and test, making sure the recordings of a given patient go to one subset. Table~\ref{tab:data-description} summarizes key statistics of the data.

\setlength\arrayrulewidth{1.0pt}
\begin{table}[h]
    \centering
    \begin{tabular}{c|ccc|ccc}
      & \multicolumn{3}{c|}{\textbf{Healthy}} & \multicolumn{3}{c}{\textbf{Neuro Injury}} \\
      & Train & Val & Test & Train & Val & Test \\
     \textbf{\# recordings} & 1360 & 247 & 238 & 92 & 40 & 45 \\
     \textbf{\# patients} & 885 & 165 & 163 & 75 & 33 & 38 \\
     \textbf{\# hours} & 10.3 & 1.9 & 2.0 & 0.8 & 0.3 & 0.3 \\
    \end{tabular}
    \caption{\cem{The description of our} neurological injury dataset.}
    \label{tab:data-description}
\end{table}

In addition, the recordings are annotated with a trigger - the primary reason for crying determined by the medical or research staff. In this study, we use a subset of three main triggers resulting in 267 recordings of discomfort, 200 hunger, and 682 pain. 

Compared to the commonly used Chillanto database~\cite{reyes2004system}, our dataset has much more patients with neurological injury (146 vs 6) and more annotated cry signals in general (14.2 vs 0.6 hours). 
In our database, cry recording is a segment of arbitrary length (a second to a few minutes). Conversely, in  Chillanto, the recordings correspond to 1-second cry expirations annotated as belonging to the healthy or sick infant. However, there is insufficient evidence to determine whether every cry expiration of a sick infant has distinct characteristics from healthy infants or if only some expirations have them. 
Furthermore, cry expirations of a single infant are generally quite similar, so if recordings are split randomly for training, testing, and validation without considering infant identities (for example, in~\cite{patil2022constant}), the resulting performance may be over-estimated.

For SSL experiments, we also use an additional 8 hours of Ubenwa unlabeled cries along with 1.6 hours available in VGGSound and about 1.4 hours collected from Freesound website\footnote{ \url{https://freesound.org} } using search query ``baby cry''.

\subsection{Baselines}

While our primary focus is on SSL pre-training and fine-tuning, we use two supervised approaches as baselines that do not rely on pre-training. 
The first system is a statistical model using ComParE 2016~\cite{schuller2016interspeech} acoustic features extracted with OpenSmile toolkit~\cite{eyben2010opensmile}. The feature set contains 6373 recording-level derivatives (mean, standard deviation, etc.) of various acoustic descriptors (MFCC, pitch, jitter, etc.). It is commonly used in computational paralinguistics and infant cry classification~\cite{parga2020defining}. The model and hyperparameters are selected using grid search and 10-fold cross-validation, maximizing the average AUC score. For this study, random forest, logistic regression, and support vector machine were considered in model selection.
The second baseline is CNN14 described in Section~\ref{sec:method} using random initialization, no pre-training and end-to-end supervised fine-tuning described in Section~\ref{sec:protocol}
The performance of baselines on neurological injury and triggers is summarized in Table~\ref{tab:baselines} for five experiments with different random seeds. 
CNN14 without pre-training does not outperform the statistical baseline, which is not surprising given that our training datasets are quite small for such a model.

\begin{table}[h]
    \centering
    \begin{tabular}{c|c|c}
     \multirow{2}{*}{\textbf{Model}}& \multicolumn{2}{c}{\textbf{AUC \% (mean and standard error)}} \\ 
     & Neuro Injury & Cry Triggers \\ \hline
     Statistical & 75.1 {\scriptsize $\pm$ 0.4} & 71.1 {\scriptsize $\pm$ 0.2} \\
     CNN14 & 74.6 {\scriptsize $\pm$ 1.7} & 59.8 {\scriptsize $\pm$ 0.9} \\
    \end{tabular}
    \caption{Baseline performance obtained without any pre-training}
    \label{tab:baselines}
\end{table}

\section{SSL experiments}
\label{sec:exp}

The main results of neurological injury and cry trigger experiments are summarized in Tables~\ref{tab:asphyxia-results} and~\ref{tab:trigger-results}. The first row \cem{in both tables} refers to supervised training with random initialization and is provided to give an idea about the model performance without pre-training.

The last three columns \cem{in Table \ref{tab:asphyxia-results} and Table~\ref{tab:trigger-results}} \cem{summarize the performances obtained after fine-tuning the network with supervised training. From left to right, the results in the tables correspond to the following: 
\begin{enumerate}
    \item Evaluation with linear probing, where a linear layer is fine-tuned on top of the frozen encoder weights (Linear).
    \item Evaluation with linear probing, with batch-norm layers  updated during fine-tuning (Linear+BN).
    \item End-to-end fine-tuning where the linear layer and whole encoder are updated (End-to-end).
\end{enumerate}
}


\setlength{\tabcolsep}{2pt}
\begin{table}[h]
    \centering
    \begin{tabular}{c|c|c|ccc}
   & \textbf{Pre-} & \textbf{SSL cry adapt.}  & 
 \multicolumn{3}{c}{\textbf{\% AUC after fine-tuning}}  \\ 
      & \textbf{training} & \textbf{Dataset} & Linear & Linear+BN & End-to-end \\
    \hline \hline
       1 & -- & --  & 60.6 {\scriptsize $\pm$ 1.5}  & 60.4 {\scriptsize $\pm$ 1.3} & 74.6 {\scriptsize $\pm$ 1.7} \\
       2 & supervised & -- & 75.5 {\scriptsize $\pm$ 0.6}  & 75.9 {\scriptsize $\pm$ 0.7}  & 80.0 {\scriptsize $\pm$ 0.7} \\
       3 & SSL & -- & 71.3 {\scriptsize $\pm$ 0.9} & 78.5 {\scriptsize $\pm$ 1.2}  &  83.9 {\scriptsize $\pm$ 0.6} \\ \hline     
       4 & SSL & train set & 78.8 {\scriptsize $\pm$ 0.5} & 78.0 {\scriptsize $\pm$ 0.7} & 80.8 {\scriptsize $\pm$ 0.8} \\
       5 & SSL & + 11h cry &  79.8 {\scriptsize $\pm$ 0.4} & 81.3 {\scriptsize $\pm$ 0.5} & 81.3 {\scriptsize $\pm$ 0.7} \\
       6 & SSL & + replayVGG &   80.8 {\scriptsize $\pm$ 0.5}  &  83.3 {\scriptsize $\pm$ 0.6} &  {\bf 85.0} {\scriptsize $\pm$ 0.9} \\
    \end{tabular}
    
    \caption{\cem{Performance of neuro injury classification under various types of pre-training and fine-tuning strategies.   \textbf{Column 1} indicates the type of pre-training. Note that for the first row, no pre-training is applied. For the second row, supervised pre-training on the VGGSound dataset is applied. The rest of the rows use SSL-based pre-training on the VGGSound. \textbf{Column 2} indicates the datasets used in \SSLtwo. Note that in rows 4-5-6, the datasets used in SSL cry adaptation are cumulative. The 4th row uses neuro injury train, the 5th adds 11h cry to the neuro injury train, and the 6th adds a replay buffer from the VGG Sound dataset to the previous datasets from rows 4-5. \textbf{Columns 3-5} show the \% AUC (with mean and standard error) obtained with different supervised fine-tuning strategies (after the SSL fine-tuning as shown in Figure \ref{fig:method}). \cut{Linear denotes standard linear probing where a linear layer is trained on a frozen encoder, Linear+BN denotes the case where batch-norm parameters are updated in addition to the linear layer. End-to-end denotes the case where the entire encoder is fine-tuned after pre-training.}} }
    \label{tab:asphyxia-results}
\end{table}

\begin{table}[h]
    \centering
    \begin{tabular}{c|c|c|ccc}
    & \textbf{Pre-} & \textbf{SSL cry adapt.}  & 
   \multicolumn{3}{c}{\textbf{\% AUC after fine-tuning}}  \\ 
    & \textbf{training} & \textbf{Dataset} & Linear & Linear+BN & End-to-end \\
    \hline \hline
      1 &  -- & --  &  57.1 {\scriptsize $\pm$ 2.6} & 60.7 {\scriptsize $\pm$ 0.5} & 59.8 {\scriptsize $\pm$ 0.9} \\
       2 & supervised  & -- & 67.9 {\scriptsize $\pm$ 1.9}  & 68.0 {\scriptsize $\pm$ 1.7}  & 68.1 {\scriptsize $\pm$ 1.6} \\
       3 & SSL & -- & 65.9 {\scriptsize $\pm$ 0.8} & 69.5 {\scriptsize $\pm$ 0.7}  & 69.0 {\scriptsize $\pm$ 0.9} \\ \hline

       4& SSL & neuro injury train & 71.7 {\scriptsize $\pm$ 0.5}  & {\bf 75.4} {\scriptsize $\pm$ 0.8}  & 72.4 {\scriptsize $\pm$ 1.4} \\
       5 & SSL & + 11h cry &   74.5 {\scriptsize $\pm$ 0.4}  & 74.7 {\scriptsize $\pm$ 0.4}    & 72.0 {\scriptsize $\pm$ 1.8}  \\
      6 & SSL & + replayVGG & 74.2 {\scriptsize $\pm$ 0.4}  &   {\bf 75.6} {\scriptsize $\pm$ 0.6}  &   74.4 {\scriptsize $\pm$ 0.7}  \\
    \end{tabular}
    \caption{\cem{Performance of cry trigger classification. We follow the same structure used in Table \ref{tab:asphyxia-results}, therefore the same caption applies.}}

    \label{tab:trigger-results}
\end{table}

The second and third rows \cem{in both tables} compare supervised and self-supervised initial pre-training with VGGSound. In these experiments, the cry database is used only for supervised fine-tuning \cem{(In other words, no-additional \SSLtwo \; stage is applied). We observe that,} while simple linear probing performs better for supervised pre-training, the self-supervised pre-training achieves better results when updating BN statistics. Also, \cem{we see that with the end-to-end fine-tuning strategy} SSL pre-training performs significantly better on neuro injury task.  

Next, \cem{in rows 4-5-6 of Table \ref{tab:asphyxia-results} and Table \ref{tab:trigger-results} we show the results when additional \SSLtwo \;stage is employed.} First, we fine-tune with SimCLR using only the neuro injury training dataset, as shown in row 4 of Table~\ref{tab:asphyxia-results}). \cem{For neuro injury, we observe that} this significantly improves AUC for linear evaluation (71.3 to 78.8), but the improvement vanishes when BN is updated (78.5 and 78.0). We hypothesize that SimCLR in this experiment performs better mostly due to a significant domain mismatch between VGGSound that a simple BN update can compensate for. \cem{For cry triggers, we observe that \SSLtwo \; in general improves the performance obtained after supervised fine-tuning. }

\cem{Next, as shown in the 5th row of Table \ref{tab:asphyxia-results} and Table \ref{tab:trigger-results}} we further double the amount of unlabeled data for SimCLR \SSLtwo \; stage. \cem{This is achieved by adding} an 8-hour portion of previously unused Ubenwa cry along with some \cem{unlabeled} cry sounds from VGGSound and freesound. \cem{In total, these \SSLtwo \; datasets amount to approximately 11 hours of recording (hence it is called 11h cry in Table \ref{tab:asphyxia-results}, and Table \ref{tab:trigger-results})}. This significantly improves the performance of linear evaluation with and without BN update \cem{for neuro injury}. 

This, however, is not the case for end-to-end fine-tuning, where the initial VGGSound pre-training results in better transfer for neuro injury (row 3 of Table~\ref{tab:asphyxia-results}). \cem{We hypothesize that the model loses its generalization properties that are useful for fine-tuning due to catastrophic forgetting \cite{castastrophicforgetting} when we further adapt the model with SSL.} 

To mitigate this forgetting effect and preserve generalization properties that seem to be important for transfer learning, we perform \SSLtwo \; using replay technique from continual learning literature~\cite{doi:10.1080/09540099550039318}. \cem{We show this on the last row of both tables.} This is done by replaying 50\% of the VGGSound dataset when \cem{applying \SSLtwo \;stage}. We, therefore, observe that \SSLtwo \;with replay performs significantly better in all evaluations for neuro injury. We also \cem{obtain} the best results on trigger classification using linear+BN evaluation with this approach.

\section{Training with subsets of \cem{labeled} data}
\label{sec:analysis}

Obtaining labeled, high-quality medical data is extremely time-intensive and expensive. Therefore, in this section, we aim to understand if a small amount of labeled data can still yield decent performance. 
Specifically, we aim to analyze how SSL cry adaptation helps in such a scenario and which supervised fine-tuning method gives the best performance. To this end, we experiment with linear+BN and end-to-end fine-tuning using randomized subsets of labeled neuro injury dataset.

\begin{figure}[h]
    \centering
    \includegraphics[width=0.45\textwidth]{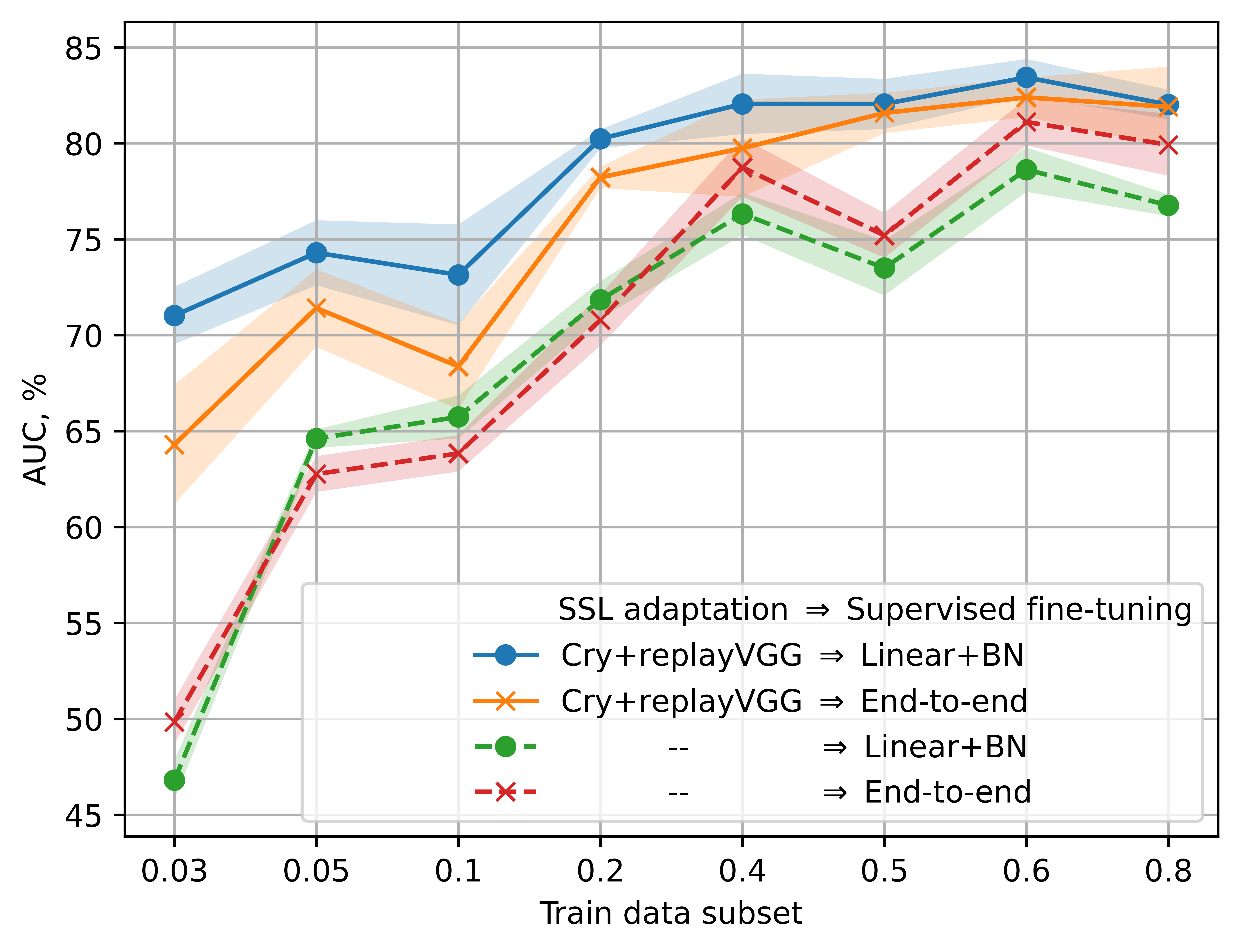}
    \caption{Performance using subsets of labeled neuro injury data in supervised fine-tuning (Linear+BN and End-to-end). Solid lines - model with SimCLR cry adaptation (last row of Table~\ref{tab:asphyxia-results}), dashed - same model without adaptation (row 3 of Table~\ref{tab:asphyxia-results}). The filled areas show the standard error of AUC from five runs with different random seeds.}
    \label{fig:auc-subsets}
\end{figure}

In Figure~\ref{fig:auc-subsets} we show results for neuro injury classification using two models: the model pre-trained with SSL on VGGSound without \SSLtwo \; stage (row 3 of Table~\ref{tab:asphyxia-results}) and the model that obtains the best performance with SSL cry adaptation (last row of Table~\ref{tab:asphyxia-results}).

The SSL cry-adapted model (solid lines) consistently outperforms the not-adapted ones (dashed lines) with a larger margin as the size of the supervised subset is reduced.
%
\cem{Another point to note is that we observe that for the not-adapted models, as the amount of labeled fine-tuning data decreases, the end-to-end fine-tuning strategy decreases more rapidly in performance. This could perhaps explain why end-to-end adaptation was performing worse than Linear+BN for cry triggers (Table~\ref{tab:trigger-results}), where less labeled data is available for fine-tuning compared to neuro injury.}

\cem{We see that, very i}nterestingly, with only 3\% (a few dozen samples) of labeled data, we can still achieve about 70\% AUC by using a cry-adapted model. 
Also, with only 20\% of data, the adapted model significantly outperforms our supervised baselines.
\cem{This showcases that SSL cry-adaptation has huge potential to obtain satisfactory model performance by only incorporating a small amount of labeled data in the supervised fine-tuning stage.} 

\section{Conclusions}
\label{sec:conclusion}

In this paper, we explored large-scale SSL pre-training for infant cry analysis, namely for detecting neurological injury and cry triggers. \cem{We observe that} SSL pre-training performs significantly better than the conventional supervised pre-training, and both perform significantly better than training from random initialization.
Furthermore, with limited annotated data, we observe that SSL adaption on cry-specific unlabeled data significantly decreases the need for labeled data in the supervised fine-tuning stage.  
\cem{We show that when we adapt the encoder through SSL using unlabeled cry data, the downstream performance for neurological injury is significantly improved. We therefore believe that, with many unlabeled cry recordings, this opens a promising research direction where it would be possible to train a classifier to detect new diseases using only a small amount of annotated cry sounds from the target population.}

\section{Acknowledgement}
We would like to thank the principal investigators from each site where cry data was collected: Dr. Uchenna Ekwochi (Enugu State University Teaching Hospital, Nigeria), Dr. Boma West and Dr. Datonye Briggs (River State University Teaching Hospital, Nigeria), Dr. Peter Ubuane (Lagos State University Teaching Hospital, Nigeria) Dr. Gabriel Variane (Santa Casa de Misericórdia de São Paulo, Brazil), and Dr. Guilherme Sant'Anna (Montreal Children's Hospital, Canada). We also thank Zhepei Wang for providing pre-trained models and his valuable advice. Finally, we are indebted to all infants, their families, and the medical and research staff at all sites for their participation and help with the acquisition of the cry data. 

\bibliographystyle{IEEEbib}
\bibliography{refs}

\end{document}